\documentclass[12pt]{article}
\usepackage{epsfig}

\newcommand{\eq}[1]{(\ref{#1})}
\newcommand{\diff}{\partial}
\newcommand{\beq}{\begin{equation}}
\newcommand{\eeq}{\end{equation}}
\newcommand{\beqn}{\begin{eqnarray}}
\newcommand{\eeqn}{\end{eqnarray}}
\newcommand{\C}{{\cal C}}
\newcommand{\arctg}{\mathop{\rm arctg}}
\newcommand{\dd}{{\mathrm d}}
\newcommand{\itep}
{~\vspace{-1.5cm}
\begin{flushright}
{\large ITEP-TH-74/99}
\end{flushright}
\vspace{1.0cm}}

\begin{document}
\baselineskip=14pt
\begin{center}

\itep

{\large\bf Confinement and short distance physics}

\vskip 1.0cm
{M.N.~Chernodub${}^a$, F.V.~Gubarev${}^{a,b}$,
M.V.~Polikarpov${}^a$, V.I.~Zakharov${}^{a,b}$}
\vskip 4mm
{\it $a$ Institute of Theoretical and Experimental Physics},\\
{\it B.Cheremushkinskaya 25, Moscow, 117259, Russia}
\vskip 4mm
{\it $b$ Max-Planck Institut f\"ur Physik},\\
{\it F\"ohringer Ring 6, 80805 M\"unchen, Germany}

\end{center}

\begin{abstract}
We consider non-perturbative effects at short distances in theories with 
confinement. The analysis is straightforward within the Abelian models in
which the confinement arises on classical level. In all cases considered
(compact $U(1)$ in $3D$ and $4D$, dual Abelian Higgs model) there are
non--perturbative contributions associated with short distances which are
due to topological defects. In QCD case, both classical and quantum effects
determine the role of the topological defects and the theoretical analysis
has not been completed so far. Generically, the topological defects would
result in $1/Q^2$ corrections going beyond the standard Operator Product
Expansion. We review existing data on the power corrections and find that
the data favor existence of the novel corrections, at least at the mass 
scale of (1-2) GeV. We indicate crucial experiments which could further
clarify the situation on the phenomenological side.
\end{abstract}

{\bf 1.} The perturbative QCD describes basic features of hard processes,
i.e. processes characterized by a large mass scale $Q$.  On the other hand,
the perturbative QCD does not encode the effects of confinement. Hence there
is a growing interest in power corrections to the parton model which might
be sensitive to the nature of confinement (for a recent review see 
\cite{review}). Moreover, in the the Abelian Higgs model the leading power
correction at short distances does reflect confinement \cite{gubarev}.
Namely, there exist short strings which are responsible for a stringy 
potential  between confined charges at vanishing distances $r\to 0$.

In this note we are exploring Abelian models with confinement in a more
regular way by including into consideration the compact three and four
dimensional $U(1)$ theory. In all the cases the confinement mechanism can
be understood classically. Moreover, we find that the confinement results
in additional terms in the static potential at short distances. How
topological defects can be manifested at short distances, is easy to 
understand on the example of the Dirac string. Naively, its energy diverges
quadratically in the ultraviolet but in compact $U(1)$ it is normalized to
zero \cite{polyakov} changing the power corrections at short distances. As
for non-Abelian theories, the Dirac strings are also allowed because of the
compactness of the corresponding $U(1)$ subgroup. In this sense, there is a
similarity between non--Abelian and Abelian cases.
However, there is an important difference as well. On the classical level,
the Dirac strings may end up with monopoles which have a
vanishing non-Abelian action. Thus, the monopoles observed within the
$U(1)$ projection of QCD (for a review see, e.g., \cite{chernodub98})
is a result of an interplay between classical and quantum effects.
In all the generality, one may say that the topological defects in QCD
are marked rather by singular potentials than by a large
non-Abelian action. Singular gauge potentials might be artifact of
the gauge fixing and it is not a priori clear whether
they can result in physical effects.
Therefore, we will turn at this point to analysis of
existing data on the power corrections. The data seem clearly favor
the novel $1/Q^2$ corrections.

\vspace{4mm}
{\bf 2.} First, we will outline very briefly the standard approach to
the power corrections which allows to account for soft non-perturbative
field configurations (for further references see, e.g., \cite{review}).
Consider first a QED example \cite{casimir}.
Namely, let an $e^{+}e^{-}$ pair be placed at distance $r$ near
the center of a conducting cage of size $L$, $L\gg r$.
Then the potential energy of the pair can be
approximated as
\beq
V_{e\bar{e}}(r)\approx-{\alpha_e\over r} ~ + ~ 
const \cdot {\alpha_e r^2\over L^3},
\qquad
L\gg r\label{cage}
\eeq
and the second term is a power correction to the Coulomb interaction.
The derivation of (\ref{cage}) is of course straightforward
classically, since the correction is
nothing else but interaction of the dipole with its images.
In the QCD case, one concludes by analogy that the heavy quark
potential at short distances looks as (for explanations
and further references see, e.g. \cite{az}):
\beq
\lim_{r\to 0}V_{Q\bar{Q}}(r)~=~-
{c_{-1}\over r} ~ + ~ const \cdot \Lambda^3_{QCD} \; r^2,
\label{potential}
\eeq
where $c_{-1}$ is calculable perturbatively as a series in $\alpha_s$.
Note the absence of a linear correction to the potential at
short distances.

On the other hand, Eq. (\ref{cage}) can be derived also in terms of
one-photon exchange. The power correction is related then to
a change in modes of the electromagnetic field
confined in the cage as compared to the case of the infinite space.
The change is of order unit at frequencies $\omega\sim 1/L$.
Similarly, the logic behind Eq. (\ref{potential}) is that the perturbative
gluon propagator is modified strongly by at $\omega\sim \Lambda_{QCD}^{-1}$.
In case of other processes, the relevant Feynman graphs can be more
complicated of course. The power corrections still correspond to the infrared
sensitive part of Feynman propagators which are obviously modified by
the physics of large distances. The Operator Product Expansion (OPE) allows
for a regular way to parameterize such corrections (for a review see
\cite{review}).

\vspace{4mm}
{\bf 3.} Intuitively, the power corrections in QCD could be very different
from the conducting cage case discussed above. Indeed color particles 
produced at short distances find no cage but rather build up the confining
field configuration in the course of the interaction between themselves 
and with the vacuum. The complicated space-time picture
of interaction in confining theories was studied by Gribov \cite{gribov}.
Thus, it could be instructive to analyze the effects of the confinement
at short distances in some simple models.

The first example of a theory where the OPE does not work in fact
goes back to the paper in Ref.~\cite{polyakov}.
However, since it has not been discussed in connection with the OPE, we will
explain this example in some detail. The action is that of free photons:
\beq
S~=~{1\over 4e^2}\int d^4x F^2_{\mu\nu}\label{action}
\eeq
where $F_{\mu\nu}$ is the field strength tensor,
$F_{\mu\nu}=\partial_{\mu}A_{\nu}-\partial_{\nu}A_{\mu}$.
Although the theory looks absolutely trivial, it is not the case
if one admits Dirac strings into the theory.
Naively, the energy associated with the Dirac strings is infinite:
\beq
E_{\mbox{\small Dirac string}}={1\over 8\pi}\int d^3r\,{\bf H}^2\sim l\cdot A
\left({\mbox{magnetic~flux}\over A}\right)^2\to \infty
\eeq
where $l, A$ are the length and area of the string, respectively.
Since the magnetic flux carried by the string is quantized and finite
the energy diverges quadratically in the ultraviolet, i.e. in the
limit $A\to 0$. However within the lattice regularization the action
of the string is in fact zero
because of the compactness of the $U(1)$, Ref.~\cite{polyakov}.

Now, the Dirac strings may end up with monopoles. The action
associated with the monopoles diverges in ultraviolet,
\beq
\int {d^3r\over 8\pi}{\bf H}^2~\sim~{1\over e^2a}\label{uvdivergence}
\eeq
where $a$ is a (small) spatial cut off. If the length of a closed monopole
trajectory is L, then the suppression of such a configuration due to a
non-vanishing action is of order
\beq
e^{-S}~\sim~\exp~(-const~L/e^2)\,.
\eeq
On the other hand, there are different ways to organize a loop of length $L$.
This is the entropy factor. It is known to grow exponentially with $L$ as
$\sim \exp(~const'L~)$. At some $e^2_{crit}\sim 1$ there is a phase
transition to the monopole condensation.

The potential between external test charges is Coulombic at all the
distances for $e^2<e^2_{crit}$ and linear for $e^2>e^2_{crit}$.
Since there are no perturbative graphs at all in the theory with the action
(\ref{action}) this phenomenon clearly goes beyond the OPE. In this case,
however, the violation of the OPE is too strong. Indeed, the lattice spacing
$a$ is the only dimensional parameter of the problem. As a result Coulomb
potential is not simply modified by linear corrections
but rather eliminated for $e^2>e_{crit}^2$ at all the distances.

\vspace{4mm}
{\bf 4.} Consider next the Dual Abelian Higgs Model with the action
\beq
\label{AHM_action} S= \int d^4x \left\{ \frac{1}{4g^2}
F^2_{\mu\nu} + \frac{1}{2} |(\diff - i A)\Phi|^2 + \frac{1}{4} \lambda
(|\Phi|^2-\eta^2)^2 \right\},
\eeq
here $g$ is the magnetic charge,
$F_{\mu\nu}\equiv\diff_{\mu}A_{\nu}-\diff_{\nu}A_{\mu}$. The gauge boson and
the Higgs are massive, $m^2_V=g^2\eta^2$, $m_H^2= 2 \lambda \eta^2$.
There is a well known Abrikosov-Nielsen-Olesen (ANO) solution to
the corresponding equations of motion. The dual ANO string may end up with
electric charges. As a result, the potential
for a test charge--anti-charge pair grows linearly at large  distances:
\beq
V(r) ~= ~\sigma_\infty r\;, \qquad r \to \infty.
\eeq
Note that there is a Dirac string resting along the axis of the ANO string
connecting monopoles and its energy is still normalized to zero.

An amusing effect occurs if one goes to distances much smaller than
the characteristic mass scales $m_{V,H}^{-1}$. Then the ANO string
is peeled off and one deals with a naked (dual) Dirac string.
The manifestation of the string is that the Higgs field has to vanish
along a line connecting the external charges. Otherwise,
the energy of the Dirac string would jump to infinity anew.

As a result of the boundary condition that $\Phi$ vanishes on a line
connecting the charges the potential contains a stringy piece
at short distances \cite{gubarev}:
\beq
\lim_{r\to 0}{V(r)}~=~{c_{-1}\over r}~+~\sigma_0\cdot r~.
\eeq
The string tension $\sigma_0$ smoothly depends on the ratio $m_H/m_V$.
In the Bogomol'ny limit ($m_H = m_V$) which is favored
by the fits of the lattice simulations \cite{ilgenfritz} the string tension
\beq
\sigma_0~\approx~\sigma_\infty,
\eeq
i.e. the effective string tension numerically is the same at all distances.

\vspace{4mm}
{\bf 5.} Consider now $3D$ compact electrodynamics.
As is well known \cite{U1}, the charge--anti-charge potential is then linear
at large $r$. Below we consider the string tension $\sigma_0$ at small
distances and show that it has a non-analytical piece.

As usual, it is convenient to perform the duality transformation, and work
with the corresponding Sine-Gordon theory. The expectation value of the
Wilson loop in dual variables is:
\beq\label{Schi-1}
W = \frac{1}{\cal Z} \int {\cal D}\chi e^{- S(\chi,\eta_\C)},
\eeq
where
\beq \label{Schi}
S(\chi,\eta_\C) = \left({e\over 2\pi}\right)^2
\int\, d^3 x
\left\{
{1\over 2}(\vec{\diff} \chi)^2 + m_D^2 (1-\cos[\chi-\eta_\C])
\right\},
\eeq
$m_D$ is the Debye mass and $S(\chi,0)$ is the action of the model. If
static charge and anti-charge are placed at the points $(-R/2,0)$ and
$(R/2,0)$ in the $x_1, x_2$ plane ($x_3$ is the time axis), then
\beq \label{etaC}
\eta_\C = \arctg[{x_2 \over x_1-R/2}] - \arctg[{x_2 \over x_1+R/2}],
\qquad
-\pi \leq \eta_\C \leq \pi.
\eeq

Below we present the results of the numerical calculations of the
dimensionless string tension,
\beqn
\sigma = \partial E/ \partial (m_D R)\,\, , \label{sigma}\\
E = \int d^2 x
\left\{
{1\over 2}(\vec{\diff} \chi)^2 + m_D^2 (1-\cos[\chi-\eta_\C])
\right\}\,\, . \label{energy}
\eeqn
Note that the energy $E$ is measured in the units of the dimensional factor
$(\frac{e}{2\pi})^2$ ({\it cf.} \eq{Schi}). Variation of functional
\eq{energy} leads to the equation of motion
$\Delta \chi = m^2_D \sin[\chi - \eta_\C]$. For finite $R$ we can solve this
equation numerically. The energy $E$ versus $m_D R$ is shown on
Fig.1(a). At large separations between the charges ($m_D R \gg 1$) it tends
to the asymptotic linear behavior $E = 8 m_D R$ which can be obtained also
analytically \cite{U1}.

At small distances there is a contributions of Yukawa-type to the
energy (\ref{energy}), which should be extracted explicitly. Note that in
course of rewriting original $3D$ compact electrodynamics in the 
form (\ref{Schi-1}-\ref{Schi}) the Coulomb potential was already subtracted,
so that (\ref{energy}) contains Yukawa-like piece without singularity
at $R=0$. It is not difficult to find the corresponding coefficient:
\beq
\label{energy-1}
E = E^{string} - 2\pi ( K_0[m_DR]+\ln[m_D R] )
\eeq
where $K_0(x)$ is the modified Bessel function and $E^{string}$ is the energy
of the charge--anti-charge pair which is only due to the string formation.
The corresponding string tension
\beq\label{sigma-1}
\sigma^{string} = \sigma + 2\pi ( - K_1[m_DR]+ { 1\over m_D R} )
\eeq
is shown on Fig.1(b).  We found that the best fit of numerical data
for small values of $m_D R$ is by the function
$\sigma^{string} = const \cdot (m_DR)^\nu$ which gives $\nu \approx 0.6$.

\begin{figure}
 \begin{minipage}{12.5cm} \begin{center}
 \epsfig{file=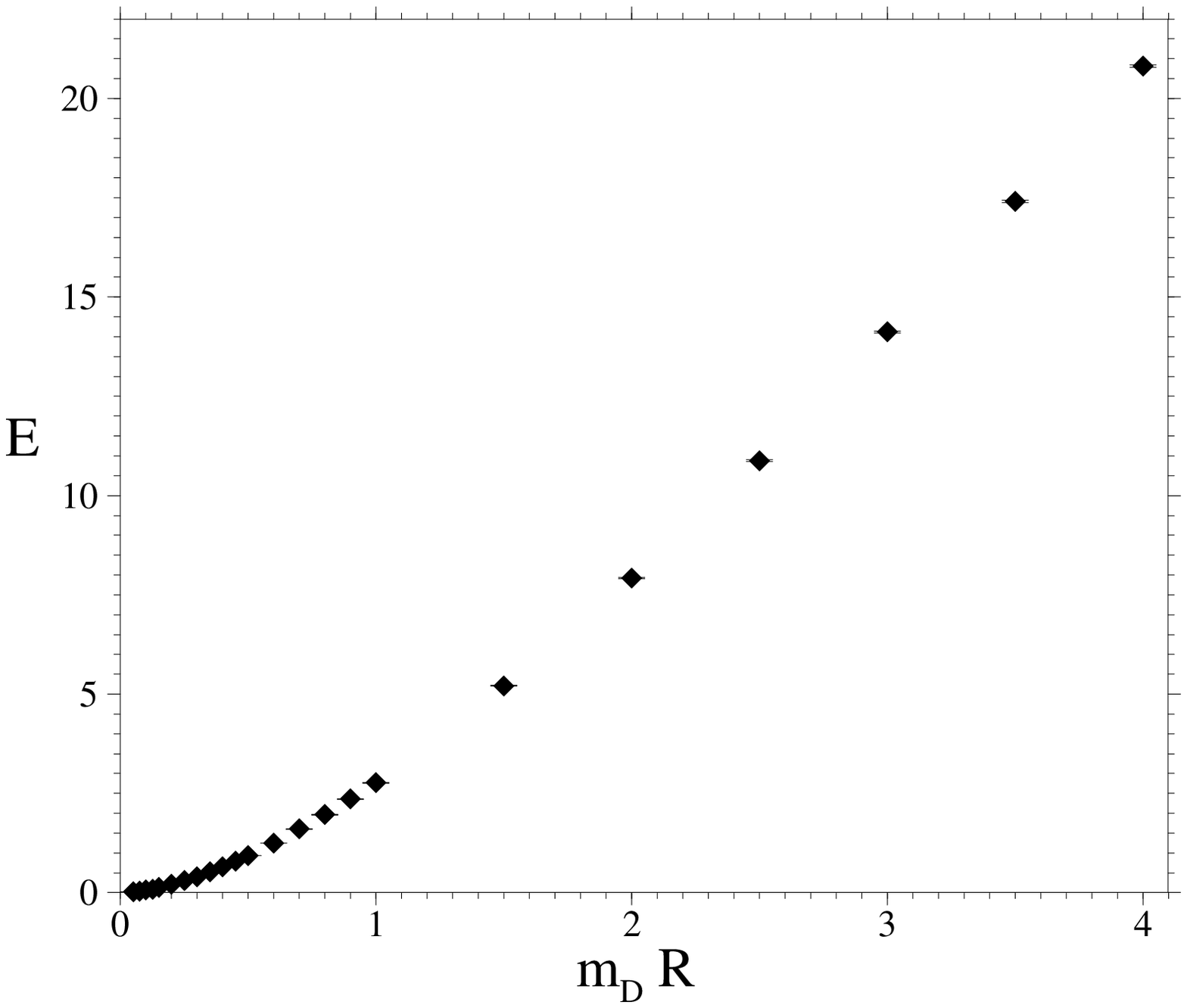,width=5.5cm,height=5.0cm} \hspace{6mm}
 \epsfig{file=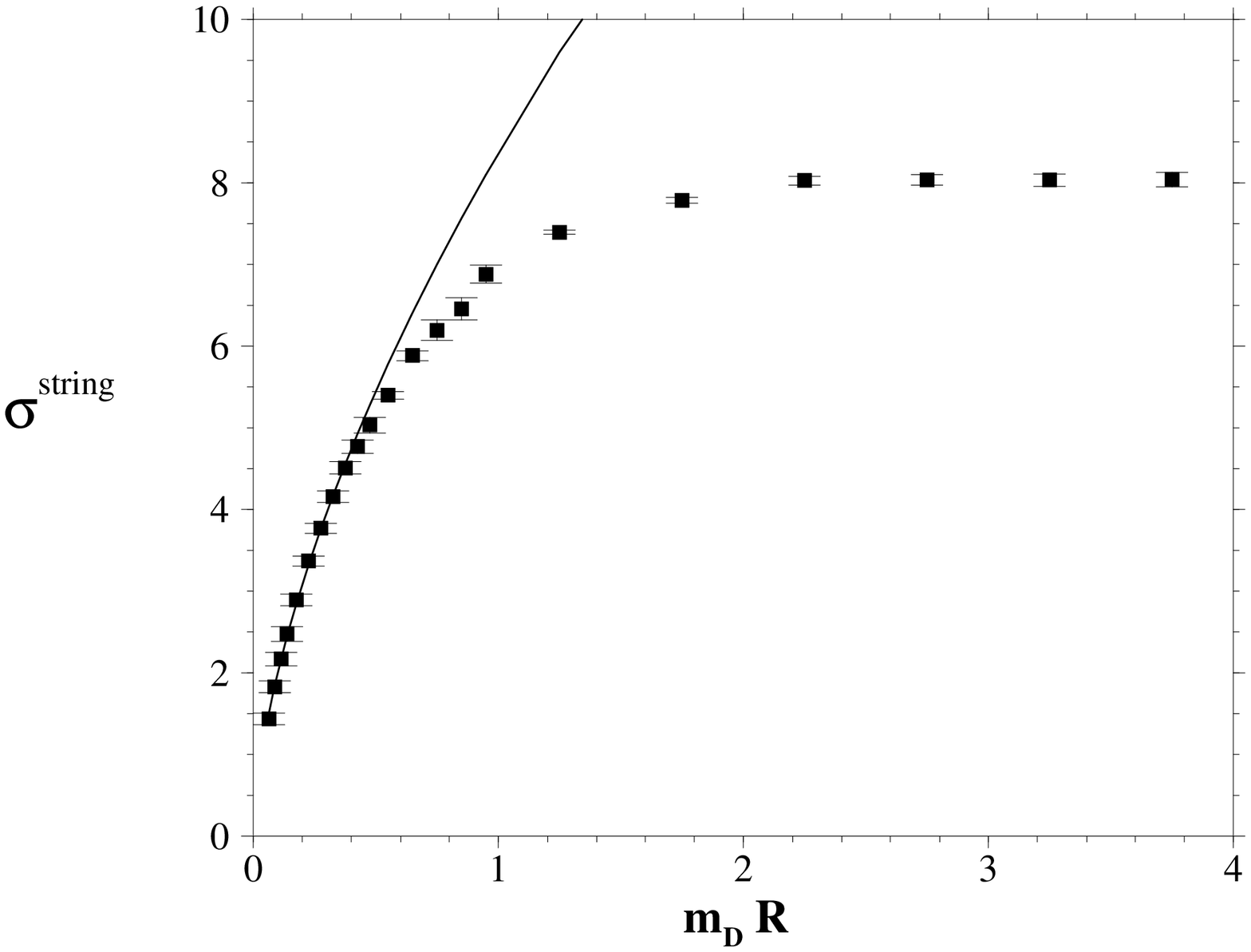,width=5.5cm,height=5.0cm}\\ (a) \hspace{5.5cm}
 (b)\\ \end{center} \end{minipage} \caption{
(a) The dimensionless string tension (\ref{sigma}) of the 
charge -- anti-charge separated by the distance $R$  in 3D compact U(1)
theory; (b) The string tension $\sigma^{string}$ (\ref{sigma-1}) as a
function of $m_D R$ with corresponding best fitting function (see text).}
\label{1}
\end{figure}

Thus the non-analytical potential associated with small distances
is softer than in the case of the Abelian Higgs model.
The source of the non-analyticity is the behavior of the function
$\eta_\C(x_1,x_2)$ eq.~\eq{etaC} which is singular
along the line connecting the charges, see Fig.2(a).

\vspace{4mm}
{\bf 6.} The compact electrodynamics is usually considered as the limit of
Georgi--Glashow model, when the radius of the 't~Hooft -- Polyakov
monopole tends to zero. For a non-vanishing monopole size the problem of
evaluating the potential at small distances becomes rather involved. To
avoid unnecessary complications we consider the 3D Georgi--Glashow model in
the BPS limit. The 't~Hooft -- Polyakov monopole corresponds then to the
fields:
\beqn
\Phi^a & = & \frac{x^a}{r} \, \Biggl(\frac{1}{\tanh (\mu r)} -
\frac{1}{\mu r}\Biggr)\,,\\
A^a_i & = & - \varepsilon^{aic} \frac{x^a}{r} \, \Biggl(\frac{1}{r} -
\frac{\mu}{\sinh(\mu r)} \Biggr)\, , \quad A^a_0 = 0\,.
\eeqn
The contribution of this monopole to the {\it full non-Abelian} Wilson loop $W$
can be calculated analytically. If the static charges are placed at points
$\pm \vec{R}/2$ in the $(x_1,x_2)$ plane the result is:
\beq \label{Wmon}
W({\vec b}_1,{\vec b}_2,\mu) = \cos h(\mu b_1) \, \cos h(\mu b_2) +
\frac{({\vec b}_1 \cdot {\vec b}_2)}{b_1\,b_2}
\sin h(\mu b_1) \, \sin h(\mu b_2)\,,
\label{WLeq}
\eeq
here ${\vec b}_{1,2} = \vec{x}_0 \pm {\vec R} \slash 2$,
$b_k = |{\vec b}_k|$, $\vec{x}_0$ is the center of the 't~Hooft -- Polyakov
monopole and
\beq
h(x) = {\pi \over 2} -
{x\over 2} \, \int\limits^{+\infty}_{- \infty}
{\dd \zeta \over  
\sqrt{\rule{0mm}{0.8\bigskipamount}x^2 + \zeta^2} \;
\sinh \sqrt{\rule{0mm}{0.8\bigskipamount}x^2 + \zeta^2}}\, .
\eeq
One way to represent \eq{Wmon} in terms of the function
$\eta_\C$ introduced earlier is:
\beqn \label{etaW}
\eta_\C(x_0,R,\mu) =
{\mathrm {sign}}(y) \, \arccos W({\vec b}_1,{\vec b}_2,\mu)\, .
\eeqn
In the limit $\mu R \to \infty$ $W({\vec b}_1,{\vec b}_2,\mu) \to \cos
\eta_\C$ and $\eta_\C(x_0,R,\mu)$ coincides with the definition \eq{etaC}.
For small $\mu R$ the function $\eta_\C$ eq.~\eq{etaW} is singular not
only between external
charges, but also outside this region (see Fig.2(b)) although the strength
of singularity gets smaller. In the limit of vanishing
$\eta_\C$ the string tension at small distances $\sigma_0$ apparently goes
to zero.

\begin{figure}
 \begin{minipage}{12.5cm} \begin{center}
 \epsfig{file=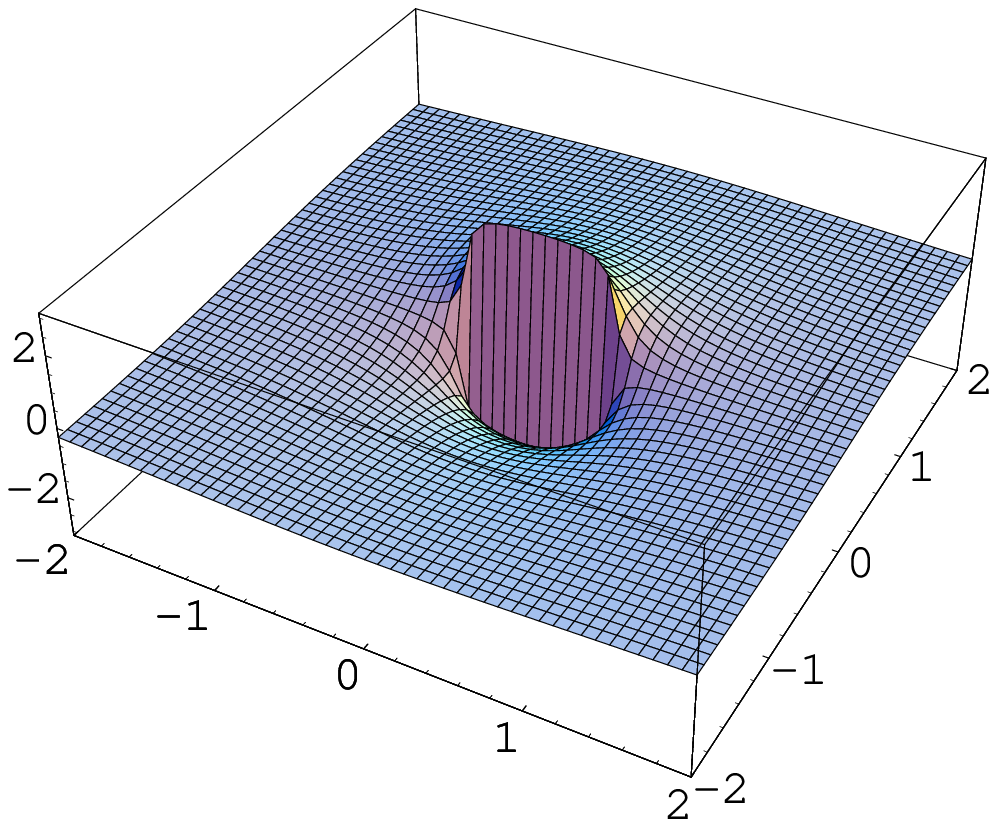,width=5.5cm,height=5.0cm} \hspace{6mm}
 \epsfig{file=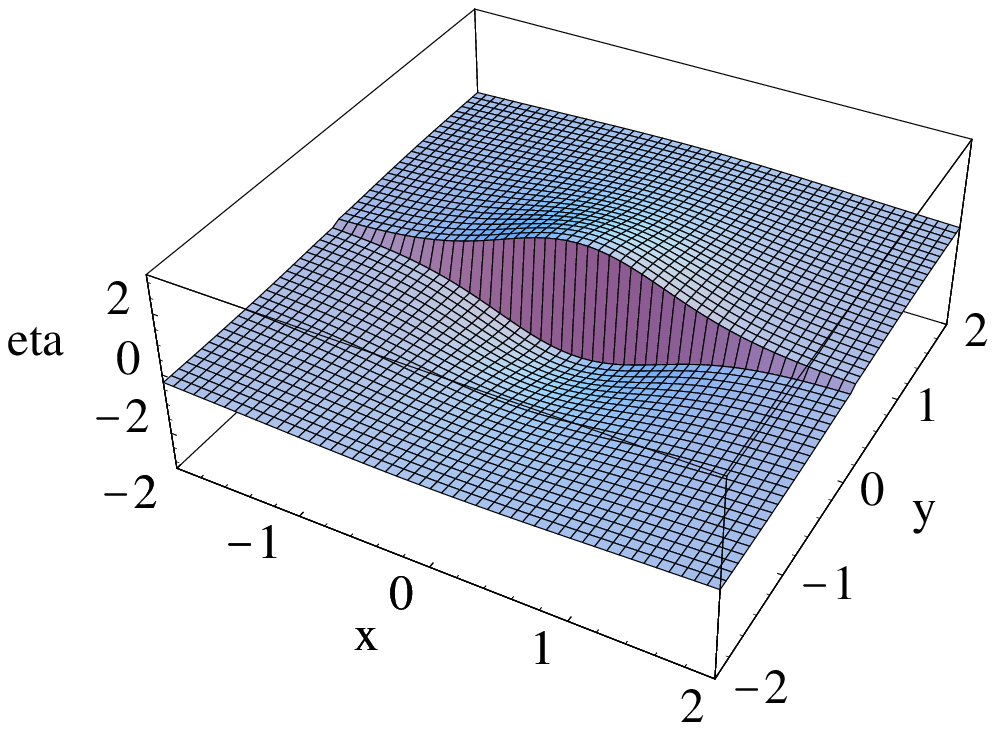,width=5.5cm,height=5.0cm}\\ (a) \hspace{5.5cm}
 (b)\\ \end{center} \end{minipage} \caption{Function $\eta_\C$ in
 compact $U(1)$ theory (a) and for extended monopoles (b),
 eq. \eq{etaW}}
\end{figure}

To summarize, it is natural to expect that in the Georgi-Glashow model
the non-analytical piece in the potential disappears at distances much 
smaller than the monopole size. Note, however, that to evaluate the 
potential consistently in this case one should have taken into account also
the modification of the interaction due to the finite size of the monopoles.

\vspace{4mm}
{\bf 7.} Knowing the physics of the Abelian models outlined above it is easy
to argue that the perturbative vacuum of QCD is not stable as well. Indeed,
let us make the lattice coarser a la Wilson until the effective coupling of
QCD would reach the value where the phase transition in the compact $U(1)$
occurs. Then the QCD perturbative vacuum is unstable against the monopole
condensation. The actual vacuum state can of course be different
from the $U(1)$ case but it cannot remain perturbative. Similar remark with
respect to formation of $Z_2$ vortices was in fact made long time 
ago \cite{mack}.

The existence of the infinitely thin topological defects
in QCD makes it close akin of the Abelian models considered above.
However, the non-Abelian nature of the interaction brings in an important
difference as well. Namely, the topological defects in QCD are marked rather
by singular potentials than by a large non-Abelian action. Consider first
the Dirac string.

Introduce to this end a potential which is a pure gauge:
\beq
A_{\mu}~=~\Omega^{-1}\partial_{\mu}\Omega\label{leer}
\eeq
and choose the matrix $\Omega$ in the form:
\beq
\Omega(x)~=~\left(
\begin{array}{cc}
\cos{\theta\over 2} & \sin{\theta\over 2} \; e^{-i\varphi}\\
\rule{0mm}{1.3\bigskipamount}
-\sin{\theta\over 2} \; e^{i\varphi} & \cos{\theta\over 2}\\
\end{array}
\right)
\eeq
where $\varphi$ and $\theta$ are azimuthal and polar angles, respectively.
Then it is straightforward to check that we generated a Dirac string directed
along the $x_3$-axis ending at $x_3=0$ and carrying the color index $a=3$.
It is quite obvious that such Abelian-like strings are allowed by
the lattice regularization of the theory.

The crucial point, however, is that the non-Abelian action associated
with the potential \eq{leer} is identical zero. On the other hand,
in its Abelian components the potential looks as a Dirac monopole,
which are known to play important role in the Abelian projection of QCD
(for a review see, e.g., \cite{chernodub98}).
Thus, there is a kind of mismatch between short- and large-distance pictures.
Namely, if one considers the lattice size $a\to 0$, then
the corresponding coupling $g(a)\to 0$ and the solution
with a zero action \eq{leer} is strongly favored at short distances.
At larger distances we are aware of the dominance of the Abelian monopoles
which have a non-zero action. The end-points of a Dirac string
still mark centers of the Abelian monopole. Thus, monopoles can be defined
as point-like objects topologically in terms of singular potentials, not
action.

Similar remarks hold in case of the $Z_2$ vortices.
Namely the $Z_2$ vortices which have a typical size
of order $\Lambda_{QCD}^{-1}$ can be defined topologically in terms of
the so called P-vortices which are infinitely thin but gauge dependent,
see \cite{Deb97} and references therein. To detect the P-vortices one uses
the gauge maximizing the sum
\beq
\sum_{l}{|Tr~U_{l}|^2}
\eeq
where $l$ runs over all the links on the lattice. The center projection is
obtained by replacing
\beq
U_{l}~\rightarrow~{\mathrm{sign}}~ (Tr~U_{l}).
\eeq
Each plaquette is marked either as $(+1)$ or $(-1)$ depending on the product
of the signs assigned to the corresponding links. The P-vortex then pierces
a plaquette with (-1). Moreover, the fraction $p$ of the total number of
plaquettes pierced by the P-vortices and of the total number of all the
plaquettes $N_T$, was found to obey numerically the scaling law
\beq
p~=~{N_{vor}\over N_T}~\sim~f(\beta )
\eeq
where the function $f(\beta)$ is such that $p$ scales like the string
tension. Assuming independence of the piercing for each plaquette
one has then for the center-projected Wilson loop $W_{cp}$:
\beq
W_{cp}=[(1-p)(+1)+p(-1)]^A~\approx~e^{-2pA}\label{tension}
\eeq
where $A$ is the number of plaquettes in the area stretched on
the Wilson loop.
Numerically, Eq. (\ref{tension}) reproduces the full string tension.

It is quite obvious that the P-vortices, since they are constructed on 
negative links, correspond in the continuum limit to singular gauge
potentials of order  $a^{-1}$.  Moreover, the large
potentials should mostly cancel if the corresponding field-strength tensors
are calculated because of the asymptotic freedom. The argumentation is 
essentially the same as outlined above for the monopoles, see, e.g.
\cite{tomboulis} and references therein.

At the moment, it is difficult to say a priori whether the topological
defects defined in terms of singular potentials can be
considered as infinitely thin from the physical point of view.
They might be gauge artifacts. Phenomenologically, using the topologically
defined point-like monopoles or infinitely thin P-vortices one can measure
non-perturbative $Q\bar{Q}$ potential at all the distances.
It is remarkable therefore that the potentials
generated both by monopoles \cite{MonSh} and P-vortices \cite{Deb97}
turn to be linear at all the distances measured:
\beq
V_{non-pert}(r)~\approx~\sigma_{\infty} r ~\mbox{  at all}~r
\label{experiment}
\eeq
Note that the Coulomb-like part is totally subtracted out through
the use of the topological defects. Moreover, the no-change in the slope
(\ref{experiment}) agrees well with the dual Abelian Higgs model
as discussed above (for alternative approaches see \cite{ImT,huber}).

The numerical observation \eq{experiment} is by no means trivial.
If it were so that only the non-Abelian action counts,
then the non-perturbative
fluctuations labeled by the Dirac strings or by P-vortices
are bulky (see discussion above) and the corresponding $Q\bar{Q}$ potentials
\eq{experiment} should have been quadratic at small $r$.
This happens, for example, in the model \cite{greensite} with finite
thickness of $Z_2$ vortices. Similarly,
if the lessons from the Georgi--Glashow model considered above apply
the finite size of the monopoles would spoil linearity of
the potential at short distances.

To summarize, direct measurements of the non-perturbative $Q\bar{Q}$
potential indicate the presence of a stringy potential at short distances.
The measurements go down to distances of order 
$(2\mbox{ GeV})^{-1}$.

\vspace{4mm}
{\bf 8.}
In view of the results (\ref{experiment}) it is interesting to reexamine
the power corrections with the question in mind, whether there is
room for novel stringy corrections. {}From the dimensional considerations
alone it is clear that the new corrections are of order $\sigma_0/Q^2$
where $Q$ is a large generic mass parameter characteristic for problem in
hand. Also, the ultraviolet renormalons in 4D indicate the same
kind of correction, see \cite{az} and references therein.
Note that unlike the case of the non-perturbative potential
discussed above, other determinations of the power
corrections ask for a subtraction of the dominating perturbative
part and this might make the results less definitive.

({\it i}) The first claim of observation of the non-standard $1/Q^2$
corrections was made in Ref.~\cite{March}. Namely, it was found that the
expectation value of the plaquette minus perturbation theory contribution
shows $1/Q^2$ behavior. On the other hand, the standard
OPE results in a $1/Q^4$ correction.

{\it (ii)} The lattice simulation \cite{bali2} do not show any change
in the slope of the full $Q\bar{Q}$ potential as the distances are
changed from the largest to the smallest ones where the Coulombic  part
becomes dominant. An explicit subtraction of the perturbative corrections
at small distances from $Q\bar{Q}$ potential in lattice gluodynamics
was performed in ref.\cite{bali3}. This procedure gives 
$\sigma_0 \approx 5 \sigma_\infty$ at very small distances.

{\it (iii)} There exist lattice measurements \cite{Fi98}
of the fine splitting of $Q\bar{Q}$ levels as function of the heavy quark
mass. The Voloshin-Leutwyler \cite{VL} picture predicts a particular pattern
of the heavy mass dependence of this splitting. Moreover, these predictions
are very different from the predictions based, say, on the Buchmuller-Tye
potential \cite{BuTy81} which adding a linear part to the Coulomb potential.
The numerical results favor the linear correction to the
potential at short distances.

({\it iv}) Analytical studies of the Bethe-Salpeter equation and comparison
of the results with the charmonium spectrum data favor a non-vanishing linear
correction to the potential at short distances \cite{BaMo99}.

({\it v}) The lattice-measured instanton density as a function of the
instanton size $\rho$ does not satisfy the standard OPE predictions that the
leading correction is of order $\rho^4$. Instead, the leading corrections is
in fact quadratic \cite{shuryak}.

({\it vi}) One of the most interesting manifestations of short strings might
be the $1/Q^2$ corrections to current correlation functions $\Pi_j(Q^2)$. It
is not possible to calculate the coefficient of front of the $1/Q^2$ terms
from first principles, however, in Ref.~\cite{CNZ99} it was suggested to
simulate this correction by a tachyonic gluon mass. On one hand, the
tachyonic mass imitates the stringy piece in the potential at short 
distances. On the other hand, it can be used in one-loop calculations
of the correlation functions. Rather unexpectedly, the use of the tachyonic
gluon mass ($m^2_g = -0.5 \mbox{ GeV}^2$) explains well
the behavior of $\Pi_j(Q^2)$ in various channels. To check the model
further, it would be very important to perform accurate calculations of 
various correlators $\Pi_j(Q^2)$ on the lattice.

\vspace{4mm}
{\bf 9.}
As seen from the points ({\it i})-({\it vi}) above, the existence of the novel
quadratic corrections is strongly supported by the data. There are, however, 
two caveats to the  statement that the novel
short-distance power corrections have been detected. On the theoretical side,
the existence of short strings has been proven only within the Abelian Higgs
model. As for the QCD itself, the analysis is so far inconclusive.
On the experimental side, the data always refer to a limited range of
distances. In particular, the linear non-perturbative potential has been
observed at distances of order of one lattice spacing which in physical units
is about $(1 \div 2\mbox{ GeV})^{-1}$. One cannot rule out that at shorter
distances the behavior of the non-perturbative power corrections
changes (see, e.g., \cite{huber,shuryak}).
Which would be a remarkable phenomenon by itself.

\vspace{4mm}
{\bf 10.} M.N.Ch. and M.I.P. acknowledge the kind hospitality of the staff of
the Max-Planck Institut f\"ur Physik (M\"unchen), where the part of
this work has been done.  Work of M.N.C., F.V.G. and M.I.P.  was
partially supported by grants RFBR 99-01230a and INTAS 96-370.

\end{document}